\documentclass[aps,prl,preprint,superscriptaddress]{revtex4}
\usepackage{graphicx}
\usepackage{color}
\usepackage{amsfonts}
\usepackage{isomath}
\usepackage{amsmath}
\usepackage{amsthm}
\usepackage{amsfonts}
\usepackage{amssymb}
\usepackage{fdsymbol}
\usepackage{braket}

\usepackage{yhmath}

\def\kbar{$\bar{\mathrm{K}}$}
\def\mbar{$\bar{\mathrm{M}}$}
\def\mbarp{\mbar$^{\prime}$}
\def\kbarp{\kbar$^{\prime}$}
\def\gbar{$\bar{\mathrm{\Gamma}}$}
\def\kgk{\kbar-\gbar-\kbarp}
\def\kg{\kbar-\gbar}

\def\mgm{\mbar-\gbar-\mbarp}
\def\WS2{WS$_2$}
\def\MoS2{MoS$_2$}{

\begin{document}

\title{Direct observation of minibands in twisted heterobilayers}
\author{ S{\o}ren Ulstrup}
\email{ulstrup@phys.au.dk}
\affiliation{Department of Physics and Astronomy, Aarhus University, 8000 Aarhus C, Denmark}
\author{ Roland J. Koch}
\affiliation{ Advanced Light Source, E. O. Lawrence Berkeley National Laboratory, Berkeley, California 94720, USA}
\author{ Simranjeet Singh}
\affiliation{Department of Physics, Carnegie Mellon University, Pittsburgh, Pennsylvania 15213, USA}
\author{ Kathleen M. McCreary}
\affiliation{Naval Research laboratory, Washington, D.C. 20375, USA}
\author{ Berend T. Jonker}
\affiliation{Naval Research laboratory, Washington, D.C. 20375, USA}
\author{Jeremy T. Robinson}
\affiliation{Naval Research laboratory, Washington, D.C. 20375, USA}
\author{ Chris Jozwiak}
\affiliation{ Advanced Light Source, E. O. Lawrence Berkeley National Laboratory, Berkeley, California 94720, USA}
\author{ Eli Rotenberg}
\affiliation{ Advanced Light Source, E. O. Lawrence Berkeley National Laboratory, Berkeley, California 94720, USA}
\author{ Aaron Bostwick}
\affiliation{ Advanced Light Source, E. O. Lawrence Berkeley National Laboratory, Berkeley, California 94720, USA} 
\author{Jyoti Katoch}
\email{jkatoch@andrew.cmu.edu}
\affiliation{Department of Physics, Carnegie Mellon University, Pittsburgh, Pennsylvania 15213, USA}
\author{Jill A. Miwa}
\email{miwa@phys.au.dk}
\affiliation{Department of Physics and Astronomy, Aarhus University, 8000 Aarhus C, Denmark}

\maketitle

\textbf{Stacking two-dimensional (2D) van der Waals materials with different interlayer atomic registry in a heterobilayer causes the formation of a long-range periodic superlattice that may bestow the heterostructure with exotic properties such as new quantum fractal states \cite{Dean:2013,Ponomarenko:2013,Hunt:2013} or superconductivity \cite{Cao:2018,Yankowitz:2019}. Recent optical measurements of transition metal dichalcogenide (TMD) heterobilayers have revealed the presence of hybridized interlayer electron-hole pair excitations at energies defined by the superlattice potential \cite{Kunstmann:2018,Alexeev:2019,Jin:2019,Seyler:2019,Tran:2019}. The corresponding quasiparticle band structure, so-called minibands, have remained elusive and no such features have been reported for heterobilayers comprised of a TMD and another type of 2D material. Here, we introduce a new X-ray capillary technology for performing micro-focused angle-resolved photoemission spectroscopy (microARPES) with a spatial resolution on the order of 1~$\mu$m, enabling us to map the momentum-dependent quasiparticle dispersion of heterobilayers consisting of graphene on WS$_2$ at variable interlayer twist angles ($\theta$). Minibands are directly observed for $\theta = 2.5^{\circ}$ in multiple mini Brillouin zones (mBZs), while they are absent for a larger twist angle of $\theta = 26.3^{\circ}$. These findings underline the possibility to control quantum states via the stacking configuration in 2D heterostructures, opening multiple new avenues for generating materials with enhanced functionality such as tunable electronic correlations \cite{Wu:2018} and tailored selection rules for optical transitions \cite{Tijerina:2019}.}

Assembling single-layer (SL) TMDs with different electronic structures in heterobilayers has emerged as a promising method for tailoring the band alignment at type-II heterojunctions \cite{Chiu:2015,Wilson:2017}, offering a means to control optical excitation and charge transfer processes at the atomic scale \cite{Hong:2014}. This approach to materials design inevitably involves joining two crystal lattices with different lattice constants and orientation. The long-range periodic pattern arising from the superposition of interlayer atomic registries produces a moir\'e superlattice. Scanning tunneling microscopy/spectroscopy (STM/STS) experiments on heterobilayers of TMDs have resolved such a moir\'e together with a local band gap modulation due to the superlattice potential \cite{Zhang:2017}. In ARPES, these superlattice effects are directly observable via the formation of minibands such as the mini Dirac cones identified in epitaxial graphene on Ir(111) \cite{Pletikosic:2009,Starodub:2011}, twisted bilayer graphene \cite{Ohta:2012} and heterostructures of graphene with hexagonal boron nitride (hBN) \cite{Wang:2016b,Wang:2016x}. Since ARPES directly probes the energy- and momentum-resolved quasiparticle excitation spectrum, these measurements provide critical information about the minibands such as dispersion, hybridization with main bands, opening of mini gaps as well as emergence of correlation effects i.e., properties that completely specify the functionality of the heterostructure.

It has so far not been possible to observe similar minibands in epitaxial SL TMDs on single-crystal metal substrates \cite{Miwa:2015} or TMDs on bilayer graphene on silicon carbide \cite{Zhang:2013,Ugeda:2014} despite visible moir\'e superlattices in the STM data \cite{Miwa:2015,Ugeda:2014,Kastl:2018}. Recent optical studies of TMD heterobilayers, however, show distinct exciton lines arising from twist angle dependent minibands \cite{Alexeev:2019,Jin:2019,Seyler:2019,Tran:2019}. It is plausible that the signature of the quasiparticle minibands may have been suppressed in these earlier ARPES measurements due to stronger TMD-substrate interactions, long-range rotational disorder, and/or quenching of the interlayer photoemission intensity as a result of the three-atomic-layer (sandwich-like) structure which essentially constitutes a SL TMD, effectively ``burying" the interface. Furthermore, the epitaxial approach in these studies did not allow for tuning of layer orientation, thereby preventing a systematic search for minibands in materials with a different moir\'e superlattice. 

\begin{figure*} [t!]
\begin{center}
\includegraphics[width=1\textwidth]{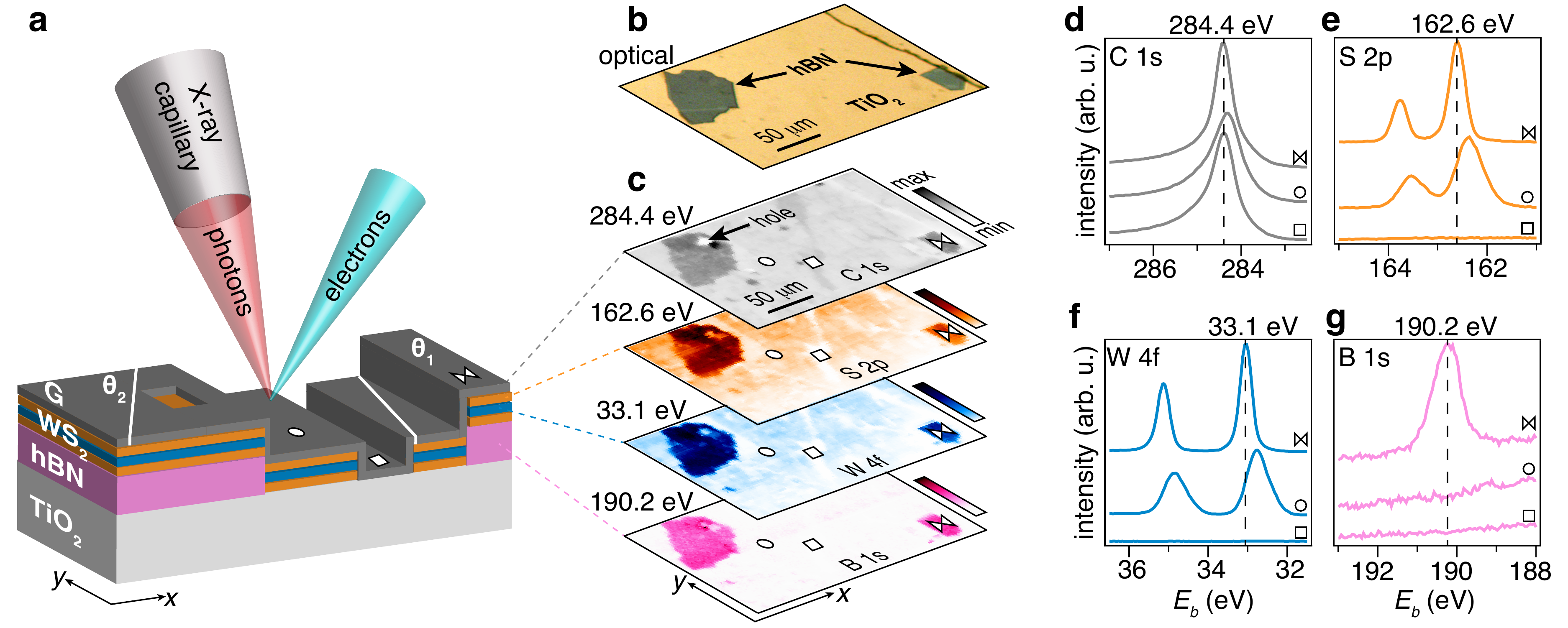}
\caption{\textbf{Elemental mapping of heterobilayer interfaces.} \textbf{a,} Model of sample and photoemission setup with a micro-focused beam achieved using an X-ray capillary. Differently rotated graphene domains are separated by white lines and annotated with twist angles $\theta_1$ and $\theta_2$. \textbf{b,} Optical microscope image of the sample. Two hBN flakes are shown via arrows. \textbf{c,} Stack of $(x,y)$-photoemission intensity maps corresponding to the given binding energies and core level peaks collected at a photon energy of 350 eV. The hBN flakes are also visible in these maps. The arrow on the C 1s map points to a hole in the graphene flake. \textbf{d-g,} Core level peaks obtained from the spots on the map in \textbf{c} labeled by corresponding symbols. The vertical dashed lines mark the binding energies used for composing the maps in \textbf{c}.}
\label{fig:1}
\end{center}
\end{figure*}

We resolve these issues by using a new microARPES approach with a spatial resolution on the order of 1 $\mu$m to measure a heterobilayer consisting of graphene, SL WS$_2$ and hBN.  Both the graphene and the SL WS$_2$  were grown by chemical vapor deposition (CVD) and sequentially transferred onto the hBN. The graphene has multiple domain orientations while the SL WS$_2$ is single domain. The resulting graphene/WS$_2$/hBN stack is supported on a TiO$_2$ wafer; see illustration in Fig. \ref{fig:1}(a) and optical microscope image of  hBN flakes on TiO$_2$ in Fig. \ref{fig:1}(b). The atomically flat interface and vanishing interlayer interactions with hBN facilitate the collection of extremely high-quality ARPES spectra as recently demonstrated for bare WS$_2$/hBN \cite{Katoch:2018}. The CVD grown graphene provides a multitude of different orientations on top of WS$_2$, giving access to domains with different twist angles ($\theta$-domains). The interface of the heterobilayer is notably situated below a single carbon layer instead of the sandwich-like structure of the SL TMD, thereby greatly enhancing the photoemission intensity from the interface compared to earlier studies. Each interface type and $\theta$-domain is measured by scanning the micro-focused beam of photons across the same area of the sample as seen in the optical microscope image in Fig. \ref{fig:1}(b), leading to photoemission from microscopic points on the sample. The focusing is achieved using an achromatic X-ray capillary, providing efficient mapping of valence band (VB) and core level binding energy regions due to a high photon flux and highly tunable photon energy range compared to conventional X-ray focusing optics such as Fresnel zone plates \cite{Sakdinawat:2010,Koch:2018b,Kastl:2019}. 

Photoemission intensity maps acquired at core level peak energies characteristic of the 2D materials in the heterostructure are shown in Fig. \ref{fig:1}(c). Each $(x,y)$-position in a map contains a measurement of the corresponding core level binding energy region as shown for the points marked by the symbols ``$\square$", ``$\bowtie$" and ``$\circ$" for C 1s in panel~(d), S 2p in panel~(e), W 4f in panel~(f) and B 1s in panel~(g). The contrasts provided by the peak amplitude, position and linewidth clearly outline the graphene/TiO$_2$ ($\square$), graphene/WS$_2$/TiO$_2$ ($\circ$) and graphene/WS$_2$/hBN/TiO$_2$ ($\bowtie$) interfaces. We also identify a hole in the transferred graphene, exposing a bare WS$_2$/hBN area as seen via an arrow in the C 1s map in panel~(c).  The spectral linewidths of W 4f, S 2p and C 1s core levels obtained on hBN (marked by the $\bowtie$) are reduced by a factor of 2 compared to those measured on the TiO$_2$ (marked by the $\circ$). Similar conclusions can be made for the linewidths of the VB spectra as discussed in further detail in the Supplementary Section 1, which we attribute to the flatness and extremely weak charge impurity scattering in hBN compared to the oxide.

\begin{figure*} [t!]
\begin{center}
\includegraphics[width=1\textwidth]{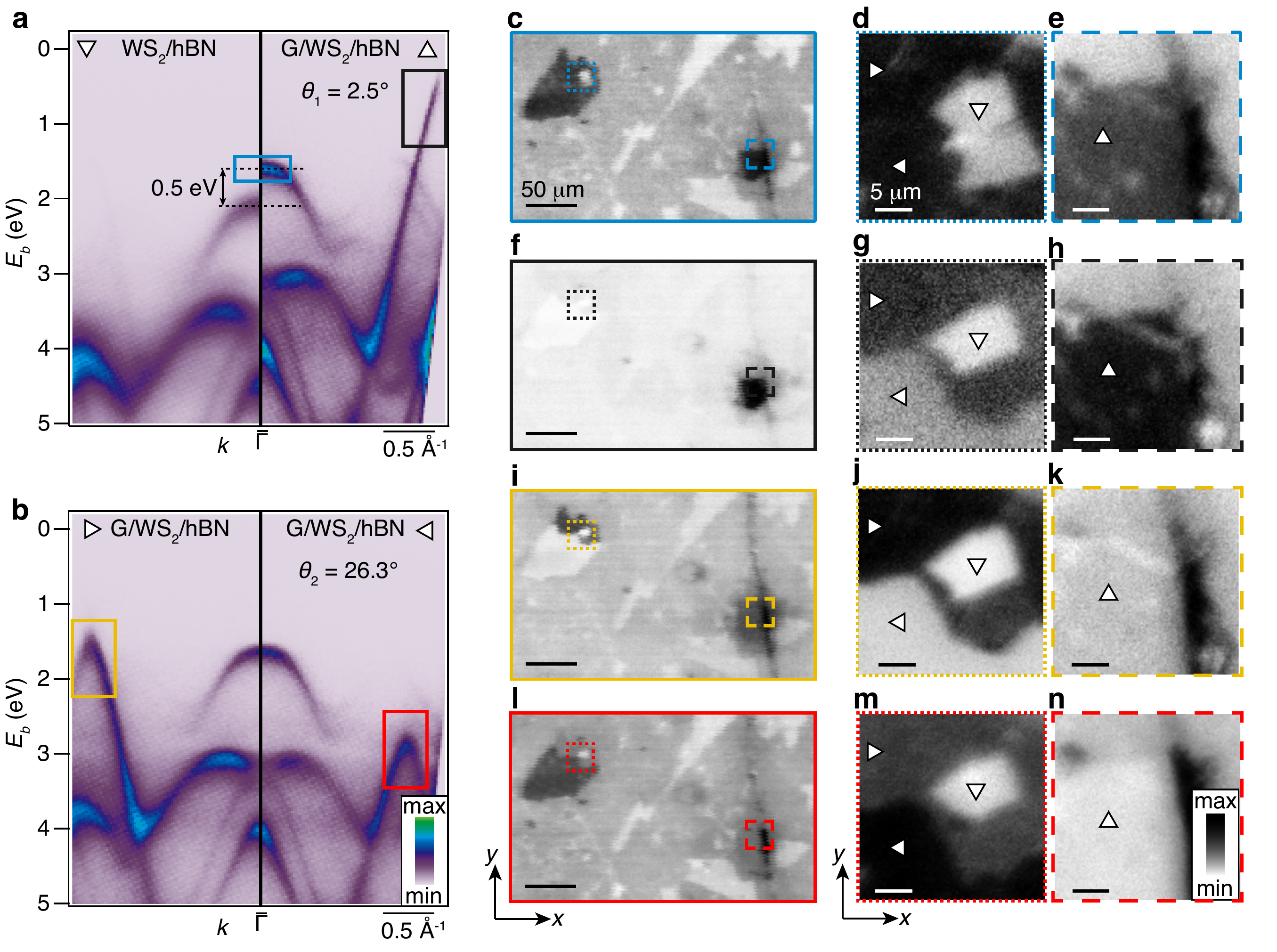}
\caption{\textbf{Band structure mapping in real space and momentum space.} \textbf{a-b,} microARPES dispersion from $\theta$-domains with the stated angles and from a bare WS$_2$/hBN area. The double-headed arrow and dashed lines in \textbf{a} indicate a binding energy shift of 0.5 eV of the WS$_2$ electronic structure on hBN in the absence of graphene. \textbf{c-n,} Maps of the $(x,y)$-dependent photoemission intensity composed from the $(E,k)$-regions marked by boxes in \textbf{a-b} with the color coding between boxes and panels given by \textbf{c-e,} blue, \textbf{f-h,} black, \textbf{i-k,} yellow and \textbf{l-n,} red. The fine scan maps in dotted and dashed panels are close-ups of the regions demarcated by dotted and dashed boxes in the coarse scan maps in \textbf{c}, \textbf{f}, \textbf{i}, and \textbf{l}. The symbols in the fine scan maps mark the spots where the correspondingly labeled $E(k)$ dispersion cuts in \textbf{a-b} were obtained. The data were obtained at a photon energy of 145 eV.} 
\label{fig:2}
\end{center}
\end{figure*}

We determine the energy- and momentum-dependent dispersion relation, $E(k)$, in the VB region, as shown in Figs. \ref{fig:2}(a)-(b). The spectra are characterized by a mix of graphene bands (see black, yellow and red boxes) and the WS$_2$ VB manifold (see blue box for the local VB maximum (VBM) at \gbar). Interestingly, we find that the graphene pins the WS$_2$ electronic structure by a rigid shift of 0.5~eV compared with the WS$_2$ states measured on a bare WS$_2$/hBN region of the sample; see double-headed arrow in Fig. \ref{fig:2}(a). The presented data are only a subset of spectra from a 4-dimensional data set containing the $(E,k,x,y)$-dependent photoemission intensity. It is furthermore possible to obtain corresponding cuts of the intensity in real space composed from a specified region of $(E,k)$-space. The results of projecting the specified electronic band structure on a $(x,y)$-dependent intensity map are shown in Figs. \ref{fig:2}(c)-(n). Panels~(c)-(e) are $(x,y)$-maps derived from the WS$_2$ local VBM at \gbar. The map in panel~(c) is a coarse scan corresponding to the field of view in Figs. \ref{fig:1}(b)-(c) while the maps in panels~(d)-(e) are fine scans of the regions demarcated by dotted and dashed boxes in panel~(c). High (low) intensity in real space allows us to identify the presence (absence) of WS$_2$ electronic states at energies and momenta determined by the blue box in panels~(a)-(b). Similarly, the $(x,y)$-maps in panels~(f)-(n) track the domains with the graphene bands given by the $(E,k)$-cuts in the black, yellow and red boxes in panels~(a)-(b). Measurements of the full 2D Brillouin zone (BZ) permit the determination of  $\theta$, as explained in the Supplementary Section 2. Here we focus on the two $\theta$-domains seen via high intensity in panels (f) and (l) (see also symbols ``$\triangle$" and ``$\triangleleft$" in Fig. \ref{fig:2}) where we find two significantly different twist angles given by $\theta_1 = (2.5 \pm 0.2)^{\circ}$ and $\theta_2 = (26.3 \pm 0.2)^{\circ}$. We also observe another orientation in panel (i) (marked by a ``$\triangleright$" in Fig. \ref{fig:2}) but this graphene/WS$_2$ interface has a $5$ $\mu$m hole (marked by a ``$\triangledown$" in Fig. \ref{fig:2}), which led to spurious features in the BZ maps we collected from this domain preventing a detailed analysis of the twist angle.

\begin{figure}  [t!]
\begin{center}
\includegraphics{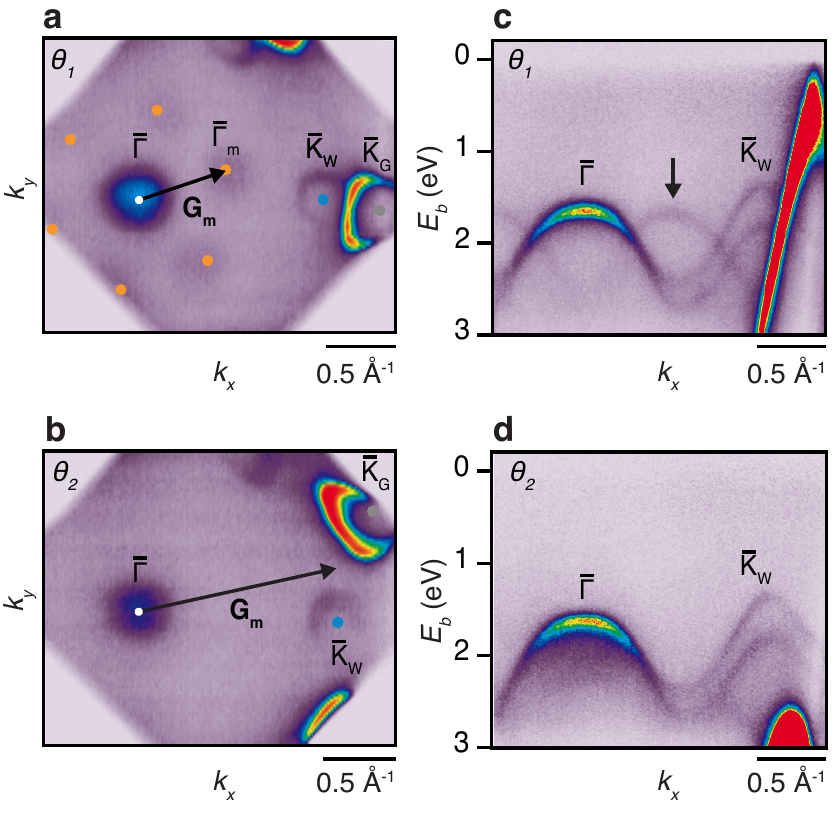}
\caption{\textbf{Direct observation of minibands.} \textbf{a,} 2D BZ map acquired on the $\theta_1$-domain at a photon energy of 70 eV.  The twist angle between the overlaying graphene and the underlying WS$_2$ is determined to be 2.5\,$^{\circ}$, as measured from the angle between the \gbar\textendash\kbar$_G$ and \gbar\textendash\kbar$_W$ directions. This twist gives rise to minibands of \gbar, marked with an orange dot and labelled as \gbar$_{m}$. \textbf{b,} Similar 2D BZ map acquired of the $\theta_2$-domain.  Here, the twist angle is 26.3\,$^{\circ}$. \textbf{c,} The dispersion for the $\theta_1$-domain along the \gbar\textendash\kbar$_W$ direction is shown with the miniband highlighted by the black arrow. \textbf{d,} Similar dispersion for the $\theta_2$-domain where the minibands are notably absent compared with the $\theta_1$-domain.}
\label{fig:3}
\end{center}
\end{figure}

Figs. \ref{fig:3}(a)-(b) map in two momentum directions, $k_x$ and $k_y$, the constant-energy contours at a binding energy of approximately 1.6~eV for the $\theta_1$- and $\theta_2$-domains. In both maps, the three most prominent features are labelled \gbar, \kbar$_W$ and \kbar$_G$; these labels correspond to the nearly circular feature of the WS$_2$ band at \gbar, the spin-split bands of WS$_2$  at \kbar$_W$, and the horseshoe-shaped arcs of the graphene Dirac cones at \kbar$_G$.  There are faint circular features surrounding \gbar\, in panel (a) that are notably absent in panel (b). These features are ascribed to minibands, labelled as \gbar$_m$, and are replicas of the WS$_2$ band at \gbar\, that arise from the periodic potential of the $\theta_1$-domain, established by the twist angle of  2.5\,$^{\circ}$ between the graphene and WS$_2$ lattices.  From the dispersion of the bands, acquired along the \gbar--\kbar$_W$ direction as shown in Fig. \ref{fig:3}(c), we see that the miniband, highlighted by the black arrow, replicates the overall shape of the SL WS$_2$ band at \gbar.  No miniband is present in the similarly acquired $(E,k)$-plot of  Fig. \ref{fig:3}(d) for the $\theta_2$-domain, where the twist between the graphene and SL WS$_2$ lattices is at a substantially larger angle of 26.3\,$^{\circ}$. As described in Supplementary Section 3, the intensity of the WS$_2$ minibands peaks in the photon energy range of 60-70~eV and appears to exhibit a strong $(E,k)$-dependence in agreement with recent theoretical predictions of the one-electron dipole matrix elements calculated for the photoemission process involving van der Waals heterostructures \cite{Amorim:2018}.

\begin{figure} 
\begin{center}
\includegraphics{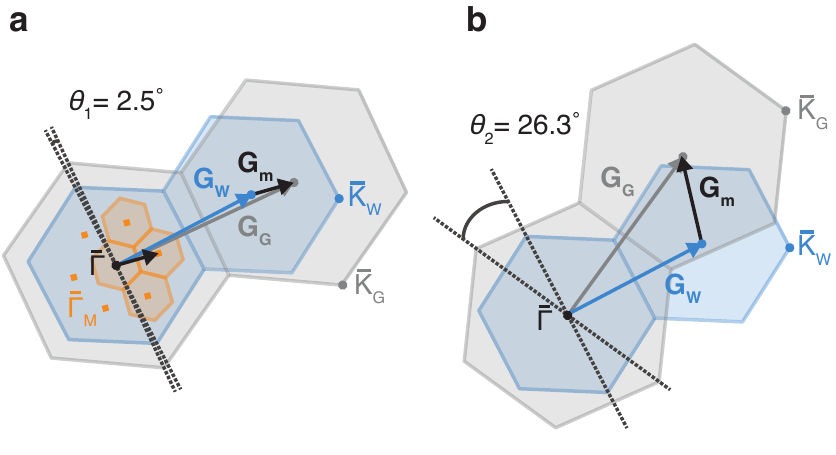}
\caption{\textbf{Construction of BZs and WS$_2$ mBZs.}  \textbf{a,} Illustration of $\theta_1$-domain with salient reciprocal lattice vectors denoted along with the formed mBZ. Two adjacent 2D BZs of graphene (grey) and WS$_2$ (blue) are overlaid with a twist angle of 2.5\,$^{\circ}$. The reciprocal lattice vectors are shown for graphene (\textbf{G}$_G$) and for WS$_2$ (\textbf{G}$_W$) that give rise to the moir\'e reciprocal lattice vector \textbf{G}$_m$ (black arrow).  The mBZs  are constructed by replicating \textbf{G}$_m$  around the \gbar\, point (orange).  \textbf{b,}  Illustration similar to the one in panel \textbf{a} but for $\theta_2$-domain.}
\label{fig:4}
\end{center}
\end{figure}

The positions of the minibands in $k$-space are determined by addition of the reciprocal lattice vectors of graphene ($\textbf{G}_G$) and WS$_2$ ($\textbf{G}_W$), as illustrated in Fig. \ref{fig:4}. Two adjacent 2D BZs of graphene (grey hexagons) are superimposed on two adjacent 2D BZs of WS$_2$ (blue hexagons) with twist angles matching those of $\theta_1$- and $\theta_2$-domains.  From the reciprocal lattice vectors, we establish the moir\'e reciprocal lattice vector ($\textbf{G}_m$) and construct the corresponding mBZs (orange hexagons) by replicating $\textbf{G}_m$ around the \gbar\, point to locate the centers (orange dots) of the surrounding mBzs.  The position of the minibands are mathematically confirmed using the following expressions for the angles and magnitude of the moir\'e vector: $\phi_{m,G(W)}= \arctan [\sin\theta/(\cos\theta - |\textbf{G}_{G(W)}|/|\textbf{G}_{W(G)}|)]$ and $|\textbf{G}_m| = (\textit{G}_{G(W)}^2 -|\textbf{G}_{G(W)}||\textbf{G}_{W(G)}|\cos\theta)/(|\textbf{G}_{G(W)}|\cos\phi_{m,G(W)})$. Indeed, we find that for the $\theta_1$-domain ($\theta_2$-domain): $|\textbf{G}_m|= 0.66$\,\AA$^{-1}$\, (1.35\,\AA$^{-1}$),  $\phi_{m,G}=8.9^{\circ}$ (49.1$^{\circ}$) for the angle between $\textbf{G}_m$ and  $\textbf{G}_G$, and $\phi_{m,W}=11.4^{\circ}$ (75.4$^{\circ}$) for the angle between  $\textbf{G}_m$ and  $\textbf{G}_W$. All magnitudes of the reciprocal lattice vectors and angles are in agreement with the microARPES data obtained for these two domains shown in Fig. \ref{fig:3} within the experimental accuracy. The minibands are absent in the case of the $\theta_2$-domain as the superlattice potential gets weaker at high twist angles leading to a reduced photoemission intensity.  This result is both consistent with theoretical calculations \cite{Amorim:2018} and  photoemission results of epitaxial graphene on Ir(111) \cite{Starodub:2011}, and with detailed theory on the origins of minibands in TMD/TMD heterobilayers  \cite{Tijerina:2019}. The observation of WS$_2$ minibands also implies the existence of mini Dirac cones replicated by the same reciprocal lattice vector in the graphene. Indeed, we find mini Dirac cones at the expected $k$-space coordinates, but their intensity is extremely faint, as shown in the Supplementary Section 4. This observation also confirms that the minibands derive from the graphene/WS$_2$ and not the WS$_2$/hBN interface. Moreover, we do not observe any superlattice features from the bare  WS$_2$/hBN interface in the hole (see the ``$\triangledown$" in Fig. \ref{fig:2}) nor in any earlier studies \cite{Katoch:2018}, possibly due to this interface being buried below the TMD sandwich-like structure. For the sake of clarity, we have omitted the hBN 2D BZ in the sketches in Fig. \ref{fig:4} as this interface does not contribute to our observations.

In summary, we have directly measured the quasiparticle band structure for a graphene/WS$_2$ heterobilayer on hBN using a new X-ray capillary technology for carrying out microARPES. In particular, we capitalize on the 1~$\mu$m spatially resolving capabilities afforded by this technique to not only determine different interlayer twist angles between the graphene and WS$_2$ but to directly measure minibands produced by the resulting superlattice potential.  The technique itself allows for a systematic search of quasiparticle band structures that emerge from moir\'e  superlattices in van der Waals heterostructures. We believe that this is a key observation that not only compliments recent optical studies that have revealed moir\'e excitons in TMD/TMD heterobilayers \cite{Alexeev:2019,Jin:2019,Seyler:2019,Tran:2019} but also demonstrates that our microARPES approach is an ideal strategy towards superlattice engineering of band structure and correlations in 2D materials. The presence of minibands for both SL WS$_2$ and graphene implies the possibility to simultaneously excite moir\'e excitons in SL WS$_2$ and generate quantum fractal states in the neighboring graphene, all within the same material stack. We therefore expect that our results will inspire further experimental and theoretical studies of the interplay of these effects in multifunctional heterobilayer materials.

\section{Methods}
\textbf{Graphene growth on Copper.}  Graphene films were grown using low-pressure CVD with flowing H$_2$ and CH$_4$ gases \cite{Zhu:2011}. Before growth, the copper foils were electrochemically polished to improve surface cleanliness and remove oxides. These copper foils were then folded into `packets' \cite{Li:2011}, loaded into a quartz tube and pumped to a base pressure of 2\,mTorr. The copper substrates were heated to 1030\,$^{\circ}$C and H$_2$/CH$_4$ was introduced for 1.5 hours at a total pressure of $\approx$60\,mTorr. The substrates were then quenched by removing them from the hot zone while under vacuum.

\textbf{Single-layer WS$_2$ growth on SiO$_2$/Si.} Synthesis of SL WS$_2$ was performed at ambient pressure in a 2-inch diameter quartz tube furnace on SiO$_2$/Si substrates (275\,nm thickness of SiO$_2$). Prior to use, all SiO$_2$/Si substrates were cleaned in acetone, isopropanol, and Piranha etch then thoroughly rinsed in DI water. At the center of the furnace  a quartz boat containing ~1\,g of WO$_3$ powder was positioned. Two SiO$_2$/Si  wafers were positioned face-down, directly above the oxide precursor. The upstream wafer contained perylene-3,4,9,10-tetracarboxylic acid tetrapotassium salt (PTAS) seeding molecules, while the downstream substrate was untreated. The hexagonal PTAS molecules were carried downstream to the untreated substrate and promoted lateral growth of the TMD materials. A separate quartz boat containing sulfur powder was placed upstream, outside the furnace-heating zone. Pure argon (65\,sccm) was used as the furnace was heated to the target temperature. Upon reaching the target temperature of 825\,$^{\circ}$C, 10\,sccm H$_2$ was added to the Ar flow and maintained throughout the 10\,minute soak and subsequent cooling. 

\textbf{Heterostructure fabrication.} Bulk hBN crystals were exfoliated onto a $n$-doped TiO$_2$ substrate using scotch tape to obtain 10--30\,nm thick flakes. The TiO$_2$  substrate with exfoliated hBN flakes was annealed in ultra-high vacuum (UHV) at 150\,$^{\circ}$C for 15\,minutes to get rid of any unwanted tape residues.  Next, CVD grown single-domain SL WS$_2$ was transferred onto the cleaned hBN flakes using a thin PC film on a PDMS stamp employing a custom-built transfer tool \cite{Katoch:2018,Ulstrup:2016}. This was followed by another annealing cycle of the WS$_2$/hBN heterostructure in UHV at 150\,$^{\circ}$C for 15\,minutes. Separately, one side of the copper foil (with graphene grown on both sides) was spin coated with a PMMA layer. The CVD graphene layer on the backside (without protective PMMA layer) of the copper foil was etched away using reactive ion etching. Next, the PMMA/graphene was floated by etching away the copper using wet chemistry \cite{Patra:2012,Singh:2015}. The PMMA/graphene layer was subsequently transferred onto a clean WS$_2$/hBN stack.  The top PMMA layer was removed by immersing the graphene/WS$_2$/hBN heterostructure in acetone for 15\,minutes followed by a rinse in isopropanol. Lastly, the graphene/WS$_2$/hBN heterostructure was subjected to another annealing cycle in UHV.

\textbf{Spatially-resolved ARPES experiments.} The heterobilayer sample was shipped in air and inserted into the Microscopic and Electronic Structure Observatory (MAESTRO) UHV facility at the Advanced Light Source (ALS) in Berkeley. The sample was annealed at 600~K for 15~minutes prior to ARPES measurements. 

The micro-focused scanning of core level and ARPES spectra presented in Figs. \ref{fig:1}-\ref{fig:2} was carried out in the MAESTRO nanoARPES end-station using an X-ray capillary (Sigray Inc.) with a spatial resolution in this experiment given by $\Delta s = (1.83 \pm 0.03)$~$\mu$m, as explained in the Supplementary Section 5. We used a coarse-motion piezo scanner for spatial maps larger than 30~$\mu$m with a step-size of 1.5~$\mu$m and a fine-motion piezo flexure scanning stage for detailed maps below this range and with a scanning step of 250~nm. Due to the achromaticity of the capillary we were able to perform valence band measurements with variable photon energy in the range 60-160~eV and W 4f, S 2p, B 1s and C 1s core level measurements with a photon energy of 350~eV without apparent loss of spatial resolution. The high efficiency of the capillary enabled acquisition of 4D datasets of the $(E,k,x,y)$-dependent intensity in $\approx$40~minutes. The $(E,k_x,k_y)$-dependent data presented in Fig. \ref{fig:3} were collected using an electron analyzer equipped with custom-made deflectors in the MAESTRO microARPES end-station, such that the sample could be held fixed during BZ mapping. These measurements were acquired with a photon beam focused to a spot size of $\approx$10~$\mu$m using Kirkpatrick-Baez (KB) optics, enabling BZ mapping on a time scale of $\approx$10~minutes for our flakes. The Supplementary Section 1 demonstrates that high quality spectra could be collected from $\theta_1$- and $\theta_2$-domains using the larger beam on the relevant areas identified by the high spatial resolution maps in Figs. \ref{fig:1}-\ref{fig:2}. All data were obtained using hemispherical Scienta R4000 electron analyzers with the energy- and momentum-resolution set at 40~meV and 0.01~\AA$^{-1}$, respectively. The sample was held at room temperature during all the measurements.

\section{Acknowledgement}
S. U. and J. A. M. acknowledge financial support from VILLUM FONDEN (Grant. No. 15375), from the Danish Council for Independent Research, Natural Sciences under the Sapere Aude program (Grant No. DFF-6108-00409) and from Aarhus University Research Foundation.  R. J. K. is supported by a fellowship within the Postdoc-Program of the German Academic Exchange Service (DAAD).  S.S. and J.K. acknowledge the start-up funds from Carnegie Mellon University. J.K also acknowledges partial support from the Center for Emergent materials: an NSF MRSEC under award number DMR-1420451.The Advanced Light Source is supported by the Director, Office of Science, Office of Basic Energy Sciences, of the U.S. Department of Energy under Contract No. DE-AC02-05CH11231. This work was supported by IBS-R009-D1. The work at NRL was supported by core programs and the Nanoscience Institute.

\section{Author information}
The authors declare that they have no competing financial interests.
Supplementary Information accompanies this paper.
Correspondence and requests for materials should be addressed to S. U. (ulstrup@phys.au.dk), J. K. (jkatoch@andrew.cmu.edu) or J. A. M. (miwa@phys.au.dk).

\newpage

\textbf{Supporting Information}
\vspace{5mm}

\noindent\textbf{1. Overview of spatially-dependent ARPES intensity in valence band region}
\begin{figure} [b!]
\includegraphics[width=0.8\textwidth]{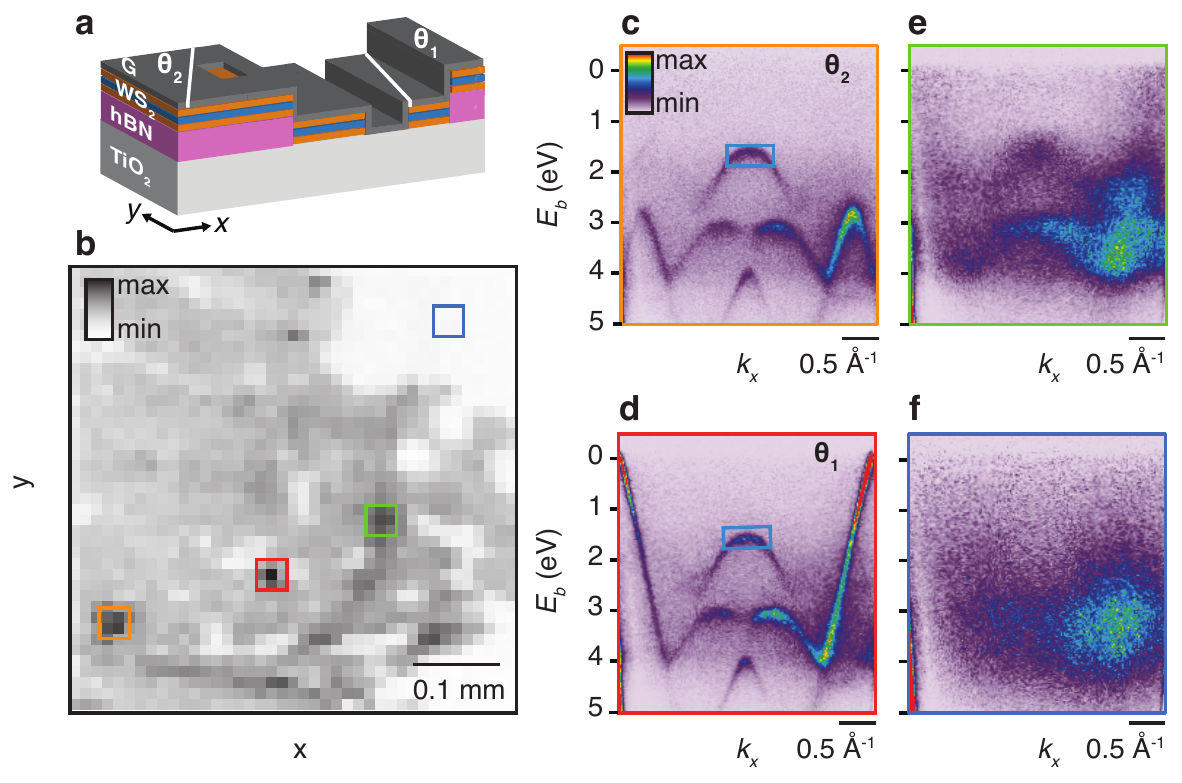}\\
\caption{\textbf{Electronic structure mapping over large sample area.} \textbf{a,} Model of interfaces present in the sample.  \textbf{b,} $(x,y)$-dependent photoemission intensity representing the majority of the sample wafer. \textbf{c-f,} Dispersion obtained from the areas in \textbf{b} marked with \textbf{c} orange, \textbf{d} red, \textbf{e} green and \textbf{f} blue boxes. The spectra were collected at a photon energy of 145 eV.}
\label{fig:1s}
\end{figure}

We collected spatial maps of the ARPES intensity using a beam with a 10~micron focus achieved via Kirkpatrick-Baez (KB) optics and conventional stepper motors for scanning in the microARPES end-station at the MAESTRO facility. The data presented here facilitates a comparison with our high resolution maps obtained with the X-ray capillary and piezo-based scanners in the nanoARPES end-station at MAESTRO (see Figs. 1-2 of the main paper). We are able to identify our heterostructures, sketched in Fig. \ref{fig:1s}(a), in the $(E,k,x,y)$-dependent intensity over a very wide area of the sample using the 10~micron beam, as demonstrated in the $(x,y)$-map in Fig. \ref{fig:1s}(b). The intensity is composed from the $(E,k)$-region marked by a box in Figs. \ref{fig:1s}(c)-(d), tracking the spatial dependence of mainly WS$_2$ states around \gbar. The orange and red boxes in panel~(b) demarcate the two hBN flakes with the graphene/WS$_2$ heterobilayers on top, characterized by the $E(k)$-dispersions in Figs. \ref{fig:1s}(c)-(d). We are able to clearly distinguish $\theta_1$- and $\theta_2$-domains with the larger beam, but detailed spatial mapping of the electronic structure requires the 1~$\mu$m beam achieved with the X-ray capillary.

We also show examples of the valence band (VB) dispersion in areas without hBN. The green box in panel (b) marks an area with a graphene/WS$_2$ heterobilayer supported directly on TiO$_2$, which leads to the broad bands seen in panel (e). We also find areas with graphene transferred on TiO$_2$, as indicated by the blue box in panel (b), which leads to the diffuse and broad graphene bands; see panel (f).\\

\newpage

\noindent\textbf{2. Determination of twist angle from BZ measurements}

\begin{figure} [h!]
\includegraphics[width=1\textwidth]{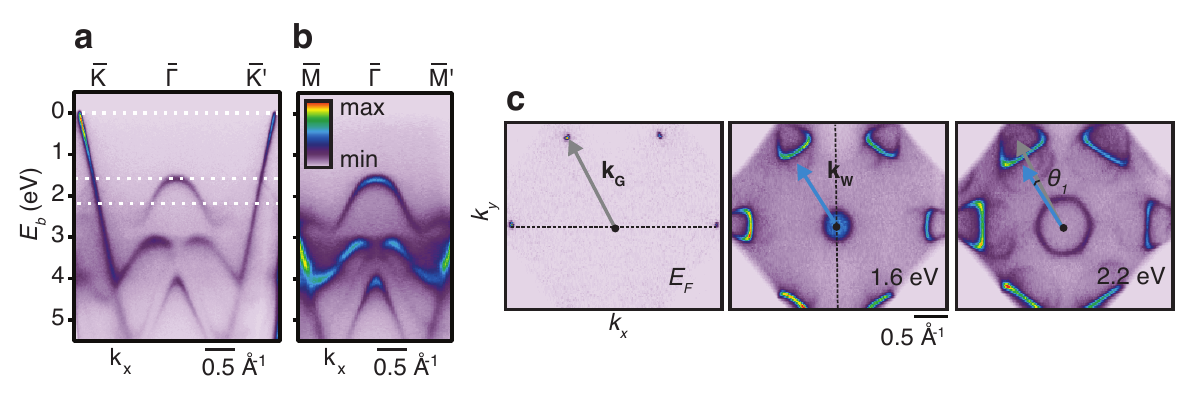}\\
\caption{\textbf{Overview of WS$_2$ and graphene BZs.} \textbf{a-b,} ARPES spectra obtained at a photon energy of 145 eV along \textbf{a} \kgk~and \textbf{b} \mgm~high symmetry directions of WS$_2$. \textbf{c,} Constant energy cuts at the given binding energies (see white dashed lines in \textbf{a}). Black dashed lines indicate high symmetry directions corresponding to the cuts in \textbf{a-b}. The vectors $\textbf{k}_W$ and $\textbf{k}_G$ describe the \kbar-points of WS$_2$ and graphene, respectively.}
 \label{fig:2s}
\end{figure}

The alignment between graphene and WS$_2$ lattices is determined using BZ scans containing multiple high symmetry points of both material BZs, as demonstrated in Fig. \ref{fig:2s}. Panels~(a)-(b) present the \kgk~and \mgm~directions of WS$_2$ while panel~(c) gives an overview of most of the 1st BZ of both graphene and WS$_2$. Such data is acquired using deflectors in the electron analyzer, thereby avoiding sample rotations which ensures that we are always measuring the same spot on the sample.

Since the graphene Fermi surface is a genuine point at \kbar~(see cut at $E_F$ in Fig. \ref{fig:2s}(c)) and the WS$_2$ \kbar~point is characterized by a circular hole pocket around the VB maximum (see cut at $E_b = 1.6$~eV in Fig. \ref{fig:2s}(c)), we can unambiguously determine the coordinates of the vectors $\textbf{k}_G$ and $\textbf{k}_W$ describing the \kbar~points of the two lattices. The angle between $\textbf{k}_G$ and $\textbf{k}_W$ corresponds to the twist angle, which is determined to be $\theta_1 = (2.5 \pm 0.2)^{\circ}$ for the example in Fig. \ref{fig:2s}(c), which corresponds to the heterobilayer in the red box in Fig. \ref{fig:1s}(b).\\

\newpage

\noindent\textbf{3. Photon energy dependence of miniband intensity}
\begin{figure} [b!]
\includegraphics[width=1\textwidth]{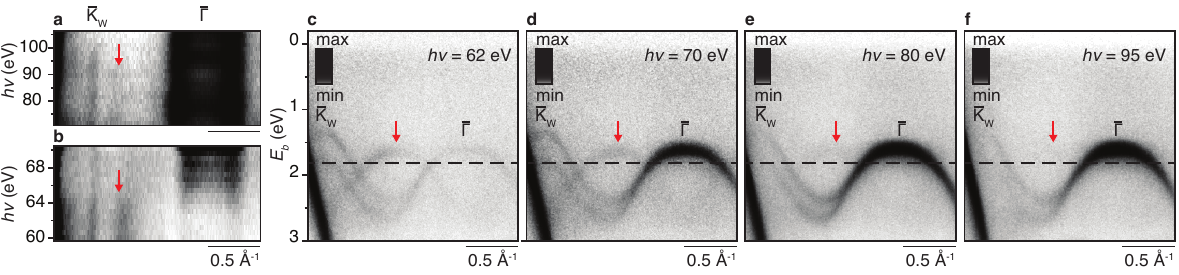}\\
\caption{\textbf{Photon energy scans.} \textbf{a-b,} Intensity of MDCs at a binding energy of approximately 1.7\,eV (see black dashed line in panels c--f) for $h\nu$ in the range \textbf{a} 70-105~eV and \textbf{b} 60-70~eV. In each $h\nu$ range we have adjusted the color scale to make the miniband intensity more clear (see red arrow).  \textbf{c-f,} WS$_2$ VB dispersion along \kg~at the given photon energies. The miniband is marked by a red arrow.}
 \label{fig:3s}
\end{figure}

The WS$_2$ miniband ARPES intensity in the $\theta_1$-domain is found to be strongly photon energy dependent due to the photoemission matrix elements, as predicted in a recent theoretical study of the photoemission intensity of van der Waals heterostructures \cite{Amorim2:2018}. 

We demonstrate this dependency in Fig. \ref{fig:3s} via photon energy scans performed along the \kg~direction of WS$_2$. The intensity of momentum distribution curve (MDC) cuts near the top of the miniband and main band at \gbar~is presented as a function of $h\nu$ in panels (a)-(b). Corresponding spectra at select photon energies are shown in panels (c)-(f). We find that the miniband intensity is comparable to that of the main band at \gbar~below a photon energy of 65 eV. We also observe that the intensity switches from one side of the miniband hole pocket to the top of the pocket (see cuts in panels (c) and (d)) as the photon energy increases. At energies higher than 80~eV the overall miniband intensity decreases (see panel (e)) until the band is no longer visible (see panel (f)). The data in Fig. 3 of the main paper were obtained at 70~eV where the main band at \gbar~is intense and the miniband intensity is uniform around the top of the band as seen in Fig. \ref{fig:3s}(d). This permits a straightforward identification of the center of the miniband hole pockets in constant energy cuts, thereby making it possible to construct the mini BZ (mBZ).

\newpage

\noindent\textbf{4. Observation of mini Dirac cones}
\begin{figure} [b!]
\includegraphics{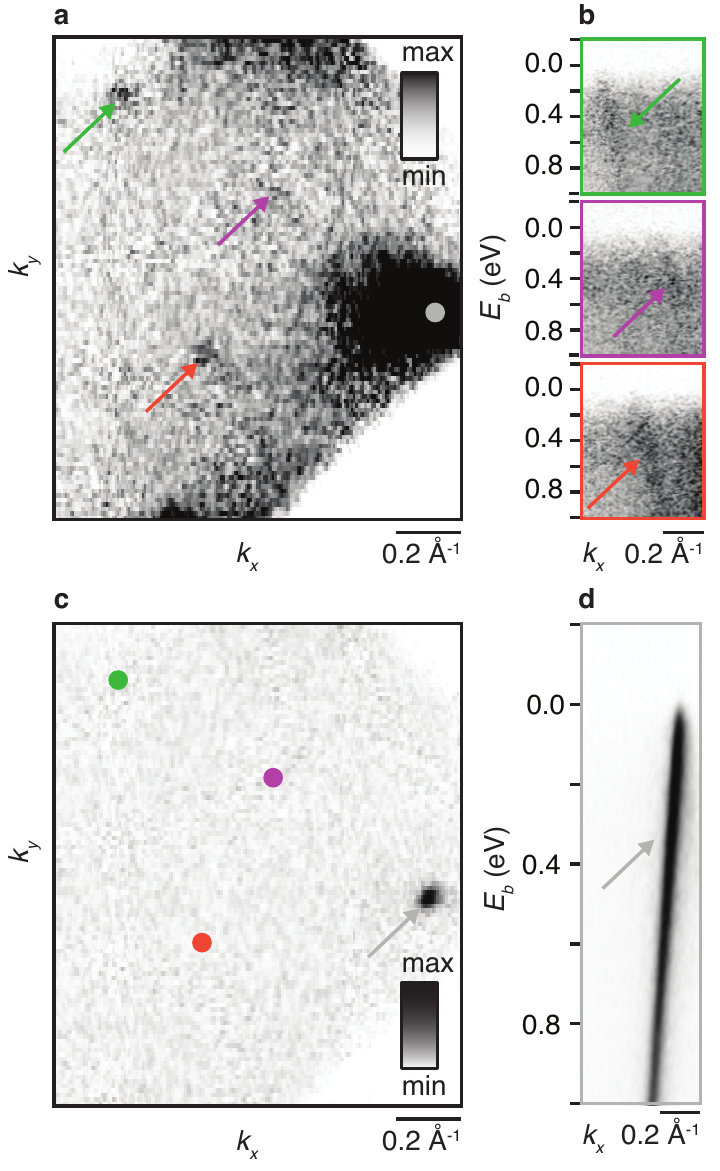}\\
\caption{\textbf{ARPES intensity from main and mini Dirac points.} \textbf{a,} Constant energy cut at approximately 0.1\,eV with the color scale enhanced to highlight mini Dirac points (see intensity at green, purple and orange arrows). The main Dirac point is marked by a light grey circle. \textbf{b,} Dispersion of mini Dirac cones obtained along the $k_x$-direction at the points marked by arrows in \textbf{a}. Arrows indicate the Dirac cones. \textbf{c,} Same cut as in \textbf{a} but with normal color scale. The circles mark the mini Dirac points while the light grey arrow marks the main Dirac point. \textbf{d,} Dispersion of the main Dirac cone obtained at the point marked by an arrow in \textbf{c}. The data is the same as shown in Figs. 3(a) and 3(c) of the main paper.}
 \label{fig:4s}
\end{figure}

The presence of WS$_2$ minibands implies the existence of similar superlattice effects in the electronic states of the graphene overlayer. Indeed, we observe very faint mini Dirac cones at the $(k_x,k_y)$-coordinates shown via arrows in the constant energy surface in Fig. \ref{fig:4s}(a) and as demonstrated in the dispersion cuts in Fig. \ref{fig:4s}(b). Note that the color scale was adjusted to enhance these features. The main Dirac cone is presented using a normal color scale in Figs. \ref{fig:4s}(c)-(d) as a reference. The data set presented here is the same as in Fig. 3 of the main paper and corresponds to the $\theta_1$-domain.

The mini Dirac points marked by purple and orange arrows in Fig. \ref{fig:4s}(a) are replicas of the main cone marked by a light grey arrow in Fig. \ref{fig:4s}(c). The green arrow in Fig. \ref{fig:4s}(a) marks a replica that belongs to a main Dirac point that is outside our measured BZ window. Given that we only observe a subset of mini Dirac points and the main Dirac cone is at the very edge of our scan, we are not able to construct the complete mBZ of graphene, but note that the locations of the mini Dirac cones we do observe are consistent with the construction in Fig. 4 of the main paper.

\newpage

\noindent\textbf{5. Estimate of X-ray capillary spatial resolution}

\begin{figure} [b!]
\includegraphics[width=0.8\textwidth]{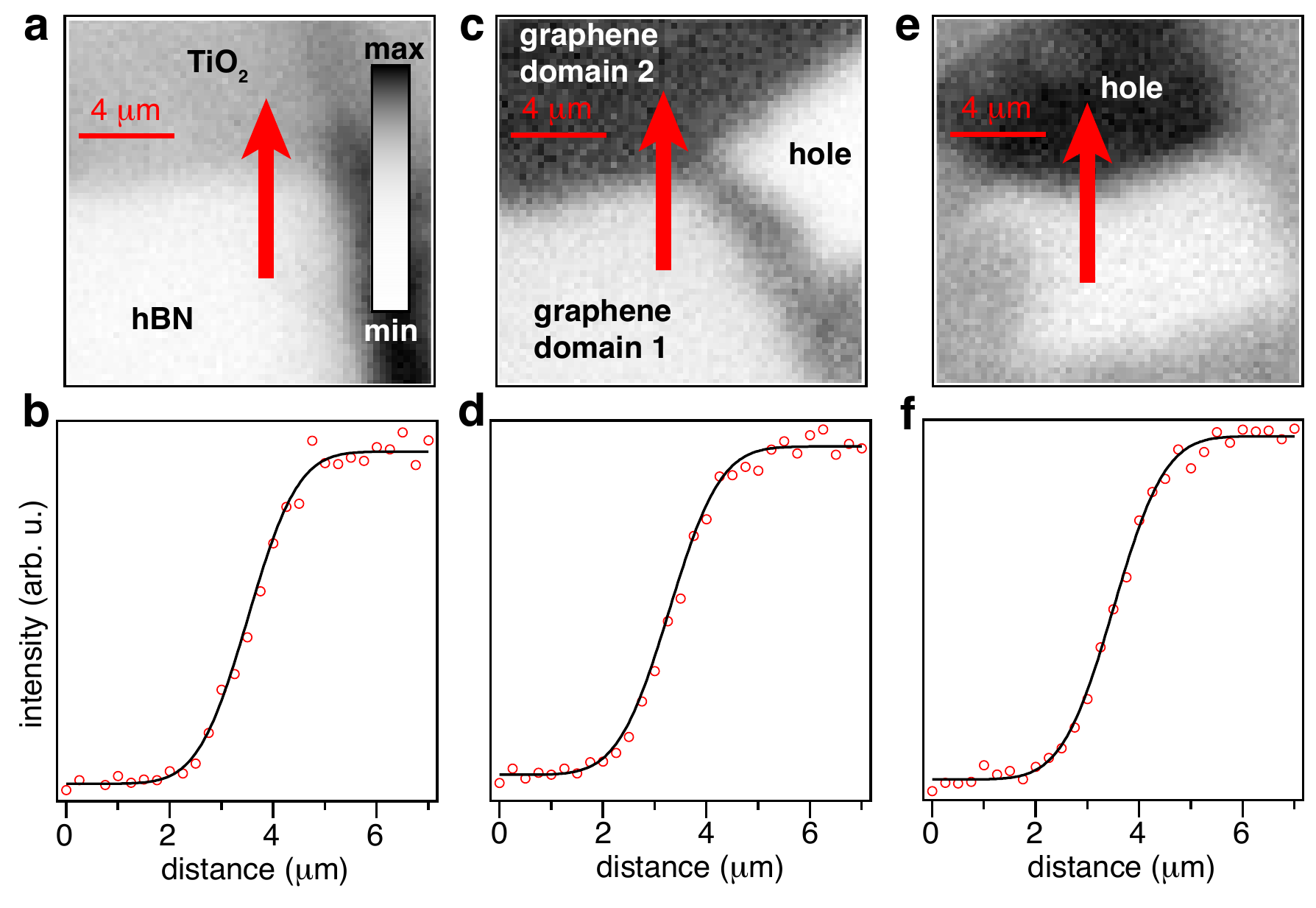}\\
\caption{\textbf{Line profile analysis.} \textbf{a,} Spatial map of photoemission intensity around the edge of a hBN flake. \textbf{b,} Line profile (open circles) acquired along the direction marked by a red arrow in \textbf{a}. The curve represents a fit to a step function broadened by a Gaussian. \textbf{c-f,} Similar analysis performed on \textbf{c-d,} a boundary between two differently rotated graphene domains and \textbf{e-f,} around a hole in the graphene.}
 \label{fig:5s}
\end{figure}

The X-ray capillary used in this experiment has been demonstrated to provide a focus on the order of 400~nm in a scanning transmission X-ray microscopy (STXM) test setup to measure a holey carbon film \cite{Koch2:2018b}. However, the actual microARPES spatial resolution will be altered by the geometry of capillary and sample required for obtaining meaningful cuts in momentum space, by the achieved focus over the measured $(x,y)$-range and by the intrinsic sharpness of the structures in the sample. In the following we estimate the spatial resolution of our experiment by assuming the sample features are much sharper than we can resolve.

Figure \ref{fig:5s}(a) shows a photoemission intensity map composed from the TiO$_2$ background intensity. This provides a sharp boundary between the hBN flake and the TiO$_2$, which we use for the line profile analysis in Fig. \ref{fig:5s}(b) where a step function broadened by a Gaussian is fitted to the profile. Similarly, in Fig. \ref{fig:5s}(c) we use the Dirac cone intensity of a graphene domain to obtain a sharp boundary with a differently rotated graphene domain and fit a line profile in Fig. \ref{fig:5s}(d). In Fig. \ref{fig:5s}(e) we use intensity from WS$_2$ bands to obtain a sharp boundary around a hole in the graphene on top of WS$_2$ and get the profile in Fig. \ref{fig:5s}(f). By analyzing several such maps composed from distinct features in our sample and using the full width at half maximum (FWHM) of the Gaussian as a measure of the spatial resolution $\Delta s$ we obtain $\Delta s = (1.83 \pm 0.03)$~$\mu$m for this experiment.

%
%
\end{document}